\documentclass[prl,twocolumn, superscriptaddress,showpacs]{revtex4-1}
\usepackage{graphicx,amsmath,amssymb,amsxtra,times,subfigure,color}
\renewcommand{\narrowtext}{\begin{multicols}{2} \global\columnwidth20.5pc}

\begin{document}

\title{Periodic Solutions and Rogue Wave Type Extended Compactons in the Nonlinear Schrodinger Equation and $\phi^{4}$ Theories}
\author{Patrick Johnson}
\author{Daniel Cole}
\author{Zohar Nussinov}
\affiliation{Department of Physics, Washington University, St. Louis, MO 63130, USA}

\date{\today}

\begin{abstract}
By employing a mapping to classical anharmonic oscillators, we explore a class of solutions to the Nonlinear Schrodinger Equation (NLSE) in 1+1 dimensions and, by extension, asymptotically in general dimensions.  We discuss a possible way for creating approximate rogue wave like solutions to the NLSE by truncating exact solutions at their nodes and stitching them with other solutions to the NLSE. The resulting waves are similar to compactons with the notable difference that they are not localized but rather extend over all of space.  We discuss rogue waves in a $\phi^4$ field theory in the context of a discretized Lagrangian and rogue wave behavior is shown to evolve into a steady state.  Due to \textit{time-reversal} invariance of this theory, the steady state found could alternatively evolve into a rogue wave giving rise to a large wave which seems to appear from nothing.
\end{abstract}


\maketitle

{\em Introduction.}
The Nonlinear Schrodinger Equation (NLSE) is a very versatile equation used in many branches of physics, dictating the behavior of wave packets in weakly nonlinear media.  It represents the evolution of optical waves in a nonlinear fiber \cite{Solli2008, Genty2010, Akhmediev1991, Dudley2008, Kasparian2009, Solli2007, Bludov2009a}, the envelope of wave packets in ocean waves in an infinitely deep ocean \cite{Akhmediev2009, Akhmediev2009a, Akhmediev2009b, Milovic2009, Ruban2006}, various biological systems, and the price of options in economics \cite{Ivancevic2010,Yan}.  One common way of solving the NLSE is through the Inverse Scattering Transform which uses the idea of Lax pairs. \cite{Tracy1988,Osborne}

A very open problem associated with the NLSE is the development of rogue waves appearing in different media \cite{Marklund2009, Efimov2008, Ruban2007, Dyachenko2005}.  Rogue waves are large waves that seem to appear from nowhere and are, at least, 2-4 times larger in amplitude than surrounding waves.  They are often preceded by a large depression referred to as a ``hole in the sea" in oceanic terms.  This is likely a nonlinear effect in most cases.

{\em Nonlinear Schrodinger Equation in 1+1 Dimensions.}
We consider the NLSE with a drift velocity and general nonlinearity
in the presence of an external potential $U$ and an additional source term
$g$.  In general dimensions, 
\begin{eqnarray}
\label{NLSE}
i \frac{\partial}{\partial t} \Psi =  [- i c \frac{\partial \Psi}{\partial x}-\epsilon \nabla^2 \Psi - \lambda F(|\Psi|^2) \Psi \nonumber
\\ + U(\vec{r},t) \Psi + g(\vec{r},t)],
\end{eqnarray}
where $\epsilon$ and $\lambda$ are tunable real parameters, $F$ is a general function of the squared modulus, and $c$ is the velocity along the drift axis (chosen to be the x-axis). 
The general ``source''  term $g$ does not make an appearance in
usual NLSE. However, it can be treated following
the transformation of the NLSE to a
mechanical problem which underlies
a central part of this work.  We will ignore $g$ and consider its effect only
in a separate later discussion. 
In 1+1 dimensions, we choose the ansatz
\begin{eqnarray}
\label{Ansatz}
\Psi(x,t) = e^{-i \omega t} u(x-ct) e^{i \delta}
\end{eqnarray}
with real $u(x) \ge 0$ and $e^{i \delta}$ an arbitrary phase.  
This gives rise to an ordinary differential equation in $u$,
\begin{eqnarray}
\label{UEquation} 
\omega u = - \epsilon \frac{d^{2} u}{dx^{2}}  - \lambda F(u^2) u +U(x,t) u.
\end{eqnarray}
Following a Galilean boost to the moving frame, $x \to (x-ct),$ 
and a subsequent interchange of space with time, 
Eq. \ref{UEquation} is the equation of motion for a nonlinear oscillator of 
spring stiffness $\omega$ subjected to an external linear force of strength $U$.  
Under this interchange,  the term $\epsilon \frac{d^{2} u}{dx^{2}}$ corresponds to the inertial term of the mass times the acceleration, and $[\omega u + \lambda F(u^2)u]$ is an effective internal forcing term.  In this form, the strength of the nonlinearity is proportional to $\lambda$.   In what follows, we first consider the case
of $U=0$ and then comment on non-zero $U$. Equation \ref{UEquation} is that of
a classical particle in an effective potential given by  \begin{eqnarray}
\label{EffectivePotentialEquation}
V_{eff}(u) = \frac{\omega}{2} u^2 + \lambda \int^{u} dv [F(v^{2}) v].
\end{eqnarray}
The energy of a classical oscillator is a constant of motion.
In this case, the corresponding ``energy''  is given
by 
\begin{eqnarray}
E = V_{eff}(u) + \frac{\epsilon}{2} (\frac{du}{dx})^{2}.
\end{eqnarray}
A particular case is that of solitons for which $u \to 0$ at large $|x-ct|$
(or large times in the corresponding mechanical problem)
and consequently $E=0$. General periodic solutions appear for
general non-zero $E$.

Rewriting Equation \ref{UEquation}, we have
\begin{eqnarray}
\label{EllipticIntegral}
\int dx = \pm \int^u \frac{dv}{\sqrt{\frac{2E}{\epsilon} -\frac{\omega}{\epsilon}v^2 - \frac{\lambda}{\epsilon}\int ^{w}\frac{d(v'^2)}{d x}F(v'^2) dv'}}.
\end{eqnarray}

\begin{figure}[htbp!]
\centering
\includegraphics[scale=0.32]{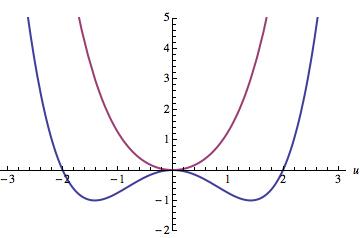}
\caption{Plot of the effective potential shown in Equation \ref{EffectivePotentialEquation} for 
a cubic non-linearity ($F =u^{2}$) for $\omega= 2$ (single minimum) and $\omega = -2$ (double minimum)
with $\lambda=1$.}
\label{EffectivePotential}
\end{figure}

For definitiveness, 
we consider the cubic nonlinearity $F(u^2)=u^2$ in our Eq.(\ref{EllipticIntegral}).  The effective potential is then  $V_{eff} = \omega u^{2}/2 + \lambda u^{4}/4$. This effective potential is plotted in Figure \ref{EffectivePotential} for 
$\omega= \pm 2$ and $\lambda=1$. In general, when $\omega<0$, we have a double minimum of 
$V_{eff}(u)$.  In such cases, when $E<0$, $u$ is restricted to be positive definite. When $\omega >0$, we have a single miniumum
at the origin ($u=0$). In all cases, as we increase the value of $E$, more values of $u$ become possible ($E \ge V_{eff}(u)$).  A transition in the form of
the effective potential (from a double well form to a single minimum)
occurs at $\omega =0$.  In passing, we note that this is similar to what occurs in Ginzburg-Landau theories 
when the coefficient of the quadratic term (in this case $\omega$) switches sign
at the transition.  We find that in the case of $F(u^{2}) = u^{2}$, 
\begin{eqnarray}
\nonumber u = -\sqrt{-\frac{\omega}{\lambda} + \frac{\sqrt{4 E \lambda + \omega^2}}{\lambda}} \times
\\ cn(\sqrt{\frac{\sqrt{4E \lambda + \omega^2}}{\epsilon}} (x-ct) ,\frac{1}{2} - \frac{\omega}{2\sqrt{4 E \lambda + \omega^2}})
\label{uexpression}
\end{eqnarray}
where $cn$ is a Jacobi Elliptic Function.  This is consistent with 
companion works \cite{PhysRevA.62.063610, PhysRevA.62.063611}.
 The Jacobi Elliptic Functions are doubly periodic  in $(x-ct)$.  In this case,
 the periodicity of the Jacobi Elliptic integral
allows a harmonic (Fourier series)
expansion of $u$  \cite{AB}.

{\em NLSE in d+1 dimensions.}
We consider the NLSE of Eq.(\ref{NLSE}) in d+1 space-time dimensions. 
With $r= |\vec{r}|$,
$\vec{r} = r \hat{n}$, and angular momentum operator, $\vec{L}$, 
in $d$ spatial dimensions, the Laplacian becomes 
$\nabla^{2} = \frac{\partial^{2}}{\partial r^{2}} + \frac{d-1}{r}  \frac{\partial}{\partial r} - \frac{\vec{L}^{2}}{r^{2}}$.  
If $\vec{L}^{2} \Psi$ is bounded in the co-moving frame at speed
$c$,  for  asymptotically large $r$ along any direction $\hat{n}$ on the unit
sphere, Eq.(\ref{NLSE}) reads
\begin{eqnarray}
\label{NLSE1}
i \frac{\partial}{\partial t} \Psi(r, \hat{n}) = [ -\epsilon \frac{\partial^{2}}{\partial r^{2}}  
\Psi(r, \hat{n}) \nonumber 
\\- \lambda F(|\Psi|^2) \Psi (r, \hat{n}) + 
U(\vec{r},t) \Psi (r, \hat{n})]. 
\end{eqnarray}
Along any direction $\hat{n}$, self-consistent solutions are enabled
by the solution of the 1+1 dimensional NLSE with the corresponding potential $U$. 
Here,``self-consistency'' is meant to imply the resulting wavefunction $\Psi$ 
has dependence on $\hat{n}$ such that $L^{2} \Psi(r, \hat{n})$ is bounded. 
The irrelevance, at large $r$, of the two terms 
$\frac{d-1}{r}  \frac{\partial \Psi}{\partial r}$ 
and $\frac{\vec{L}^{2} \Psi}{r^{2}}$ for bounded
finite wavefunctions $\Psi$ implies all large 
$r$ solutions are those of the $l=0$ (s-wave) type
with a radial dependence identical to that 1+1 dimensional
systems investigated in Eq.(\ref{EllipticIntegral}). 
A special subclass of bounded wavefunctions
of the form of Eqs.(\ref{Ansatz}, \ref{EllipticIntegral})
are those which vanish as $r \to \infty$, of pertinence 
to many usual (linear Schrodinger type) problems 
for which $\int d^{d}r |\Psi|^{2}$
is normalized. Such a vanishing of the wavefunctions
renders the higher order non-linearity irrelevant at large $r$
and reduces the problem to the usual linear Schrodinger equation. 
We consider more general waves
(idealized deep ocean type and others) for which
the wavefunctions need not vanish at
spatial infinity. In this simplest s-wave type case,
the behavior along all directions $\hat{n}$ is the same. 
In 2+1 dimensions, with an ansatz
of the form of Eq.(\ref{Ansatz}), i.e., 
$ \Psi(x,t) = e^{-i \omega t} (\sum_{m} u_{m}(r) e^{im\phi}) e^{i \delta}$
with the cylindrical coordinates $(r, \phi)$ taken
in the moving frame at velocity $c$ along the drift (x-) axis,
the NLSE in the case of cubic non-linearity reads
\begin{eqnarray}
\label{nlse2}
\omega n_{m}(r) =  [(- \frac{d^{2}}{dr^{2}} + \frac{1}{r} \frac{d}{dr}) u_{m}  \nonumber
\\ + \lambda
\sum_{m_{1},m_{2}} u_{m_{1}} u_{m_{2}} u_{m-m_{1}+m_{2}}].
\end{eqnarray}
In the large $r$ limit, under the exchange of the radial coordinate
with time,  Eq.(\ref{nlse2}), for all bounded wavefunctions with finite radial derivatives,
is that of classical oscillators 
with displacements $\{u_{m}\}$ with a uniform 
non-linear coupling $\lambda$.

{\em Noise.}
For a non-zero $U$ in Eq.(\ref{NLSE}) that depends only on
the coordinates in the drifting frame at speed $c$, 
the mechanical analogue problem of Eq.(\ref{UEquation})
along the radial direction (or, in general, in 1+1 or 2+1 dimensions) is that of an anharmonic oscillator subjected to a linear force $uU(t)$.  
Anderson localization  \cite{Anderson1958, Chabanov2000, Cheng2010, Lagendijk2009, Storzer2006} will occur for random $U(t)$. This means that
an instanton-like solution of
the mechnical problem is non-zero only in a finite interval of time.
This corresponds to a spatial localization of the wavefunction $\Psi$
over the finite interval of $|\vec{r}-ct \hat{e}_{x}|$ with $\hat{e}_{x}$ 
a unit vector along the drift direction. 

Next, we briefly consider the case of a periodic potential $U$ 
of period $a$ for $\lambda =0$. The invariance of the second order 
differential equation of motion of Eq.(\ref{UEquation}) under translations by $a$ implies that 
independent linear solutions scale as $\mu^{x/a} \Pi(x)$ with general complex $\mu$
which may be of unit modulus or real. The case with complex $|\mu|=1$ constitutes 
a mechanical analogue of
Bloch's theorem. Normalization demands that for an ideal periodic potential, 
$|\mu|=1$. For real $\mu$ with $|\mu|>1$, there is an instability (parametric resonance) 
wherein $u$ exhibits unbounded increase.  Such a case indeed occurs for a linear oscillator 
when
the potential is $U = h \cos \gamma x$ 
with $|\gamma - 2 \sqrt{\omega/\epsilon}|<h\sqrt{\omega/4\epsilon}$
wherein when coupled to the unperturbed $u$, the noise term $uU$ acts as a periodic external force with a Fourier component having a period that is close to that of the natural resonant period of the unperturbed  oscillator. This leads to large oscillations. A similar resonant 
Fourier component is generated  by
a cubic term in $u$. It is conceivable that such a situation is emulated 
for nearly periodic $U$ over some spatial range also in the presence 
of non-linearities. We speculate that noise and effects from the ocean floor along with collisions with other waves may act as effective boundary conditions on the waves similar to those discussed in 
\cite{Berry2007} which may trigger wave amplification in some cases.

{\em Source terms.}
Following the outlined prescription of converting the NLSE with
the ansatz of Eq.(\ref{Ansatz}) into a classical mechanical problem where
the spatial coordinate of the NLSE  in the moving frame
is replaced by a time coordinate, $f$ becomes
an external force. In particular, if in the 1+1 dimensional rendition 
of Eq.(\ref{NLSE}) we have $g = e^{i \omega t}  f(x-ct)$, then instead
of Eq.(\ref{UEquation}) we will have, in the moving frame at
speed $c$, $\omega u = - \epsilon \frac{d^{2} u}{dx^{2}}  - \lambda F(u^2) u 
+Uu + f(x)$. Under the interchange of space with time,
this corresponds to a classical non-linear oscillator
subjected to external forces $(Uu+f)$. The results of 
the classical analysis are then replicated. For instance, in the case 
of $U=\lambda=0$ and an oscillatory external ``force'' $f= f_{0} \cos \Omega t$,
a resonance appears when $\Omega= \sqrt{\omega/\epsilon}$ wherein 
$u \sim t \sin(\Omega t + \phi)$. The solution
of the homogeneous equation (with a vanishing $U=f=0$) is that of 
Eq.(\ref{EllipticIntegral}). Whenever the ``force'' $f$ is of the form
of the homogeneous solution, a resonance appears. In the case of
a cubic non-linearity ($\lambda \neq 0$), a resonance occurs when $f$ is of the form
of Eq.(\ref{uexpression}).

{\em Rogue Wave Type Solutions.}
We consider emulating rogue waves by spatially stitching solutions together at common
nodes. Following the formalism associated with compactons \cite{PhysRevLett.70.564}, we consider wavefunctions which are piecewise continuous with cutoffs at the nodes.  We can truncate the function, $u$, at the nodes and append on another function of our choosing with minimal error as long as the value of the function and its derivative at these points is close to zero.  This process differs from the process used to create compactons in that 
the wavefunction extends over all space and is not 
confined to a localized region. The process that we briefly describe produces an approximate solution to the NLSE (including, as a particular case, 
the linear Schrodinger equation) 
with deviations from a solution of the form of Eq.(\ref{UEquation}) only in the vicinity
of the nodes.  The initial form of these solutions can be expressed as
\begin{eqnarray}
\Psi(x,0) = \left\{ \begin{gathered}
u_1(x)  \quad \quad \quad x<x_1 \\
u_2(x) \quad x_1 < x < x_2 \\
u_3(x)  \quad \quad \quad x_2 < x\\ 
\end{gathered} \right.
\label{roguewaveequation}
\end{eqnarray}
\color{black}
where the common nodes of the function $u_2$ and $u_1$ and  $u_{3}$ are, respectively, denoted by $x_1$ and $x_2$.  In higher dimensions, $\{x_{i}\}$ denote the coordinates of planes (either 
Euclidean, spherical or other) along which
these solutions are stitched. 
Here, \{$u_{i}\}$ correspond to our solutions of the 
form of Eqs.(\ref{EllipticIntegral}, \ref{uexpression}). We note that perfect nodes where the function and its derivative are exactly zero cannot exist, because for $u$ of the form given in Eq.(\ref{uexpression}), the derivative is non-zero at all zeros of the function.
Below, we once again set $\epsilon = \lambda =1$ 
 and consider the cubic NLSE in the absence of an 
external potential.

\begin{figure}[htbp!]
\centering
\subfigure[]{
\includegraphics[scale=0.32]{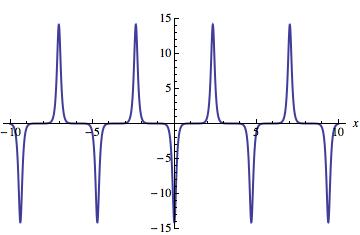}
\label{UE=10^-6omega=10^2}
}
\subfigure[]{
\includegraphics[scale= 0.32]{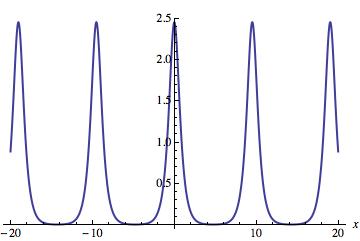}
\label{negUE=-10^5}
}
\label{RogueWavePlots}
\caption[]{Plots of the function $u$ at $t=0$ for $\lambda =1$, $\epsilon=1$, and (a) $E = 10^{-6}$, $\omega = -10^2$ and(b) $E=-10^{-5}$, $\omega=-3$.}
\end{figure}

 \begin{figure}[htbp!]
\centering
\includegraphics[scale=0.32]{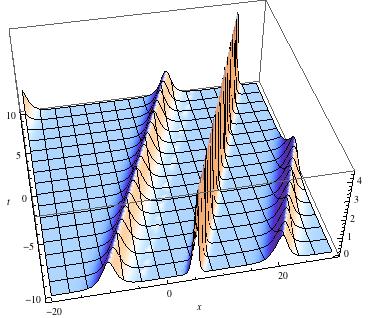}
\caption{Plot of  $|\Psi|$ of Eq. \ref{roguewaveequation} with $c=1$, $\omega_2 = -10$, $\omega_1=\omega_3 = -1$, $E=10^{-12}$, and $x_1=5.5$ and $x_2=20$.  In this plot, we shift
$(x- ct - x_0) \to x$ with $x_0 = 15$.}
\label{roguewaveplot}
\end{figure}

As $\omega$ becomes increasingly negative, the oscillations become narrower and taller as in 
Figure \ref{UE=10^-6omega=10^2}.  For negative $\omega$, negative $E$ is allowed so 
long as $E>\frac{\omega}{\lambda}$.  A form for negative $E$ is shown in 
Figure \ref{negUE=-10^5}. Although wavefunctions $\Psi$ formed by piecewise  stitching such 
solutions are special, it is clear that in order
to have transient rogue wave type phenomena, the current $\vec{j}$ cannot be 
spatially uniform (with a velocity $c$) in the rogue wave region. Far away where the current
is of uniform speed $c$, it is of the form given by
Eqs.(\ref{EllipticIntegral}, \ref{uexpression}).

A wave $u_{2}$ of large amplitude by comparison to $u_{1}$ and $u_{3}$
emulates a rogue wave. A plot of $\Psi(x,t=0)$ is provided in 
Figure \ref{roguewaveplot}. The continuity equation 
$\frac{\partial \rho}{\partial t} + \vec{\nabla} \cdot \vec{j}=0$ with the density given by 
$\rho = |\Psi|^{2}$ and the current 
$\vec{j} = i{\epsilon}(\Psi \vec{\nabla}  \Psi^{*} - \Psi^{*} \vec{\nabla} \Psi)$ applies to the NLSE.
For wavefunctions of the form of Eq.(\ref{roguewaveequation}) with real $u_{i}$,
 the current $\vec{j} =0$ at all points apart from the nodes.  
 If an interpolating form is chosen near the nodes, the current $j$ assumes
 non-zero values only in the regions close to the nodes.  If we consider the co-moving frame and define $P=\int_{x_1}^{x_2} \! \rho \, \mathrm{d}x$ where $\rho=|\psi|^2$, then integrating the continuity equation gives
\begin{eqnarray}
\label{integratedContinuity}
\frac{\mathrm{d} P}{\mathrm{d} t}=j_1-j_2
\end{eqnarray}
where $j_1$ and$ j_2$ are the currents at $x_1$ and $x_2$, respectively.  If we consider a rogue wave centered about $x=0$ extending from $x=-a$ to $x=a$ and define $P_o=\int_{X-a}^{X+a} \! \rho \, \mathrm{d}x$ with $X\gg 0$, then with $P$ the integral over the region of extension of the rogue wave, we can define the lifetime $\tau$ of the rogue wave as the time for $P$ in the region of the wave to arrive at $P_o$.  We then find that
\begin{eqnarray}
\label{lifetime}
\tau=\int_{P_o}^P \! \frac{\mathrm{d}P}{j_2-j_1}
\end{eqnarray}

Alternatively, to avoid the possible problem of slow convergence to $P_o$, we can consider the half life of the rogue wave:
\begin{eqnarray}
\tau_{1/2}=\int_{(P+P_o)/2}^P \frac{\mathrm{d}P}{j_2-j_1}
\end{eqnarray}

We can concretely see the behavior of a rogue wave type solution by evolving rogue wave initial conditions according to the linear, free-particle Schrodinger equation (LSE).  The behavior of wavefunctions under the LSE should give us qualitative understanding of the time evolution of rogue waves.

In the co-moving frame, we can write down rather general initial conditions for a rogue wave as $\psi(x,0)=\cos{x}+\alpha e^{-x^2}$, with $(1+\alpha)^2$ the amplitude of the rogue wave relative to the background, typically between $2$ and $4$. Evolving $\psi(x,0)$ according to the LSE, we find that
\begin{eqnarray}
\label{roguewavesoln}
\psi(x,t)=e^{-\frac{i}{2}t}\cos{x}+\alpha\frac{e^{\frac{-x^2}{1+2it}}}{\sqrt{1+2it}}
\end{eqnarray}
The amplitude of the rogue wave at the origin is
\begin{eqnarray}
\label{roguepeak}
|\psi(0,t)|^2=1+\frac{\alpha^2}{\sqrt{1+4t^2}}+\alpha\frac{\sqrt{1-2it}}{\sqrt{1+4t^2}}e^{\frac{i}{2}t}+c.c.
\end{eqnarray}
After an initial drop in amplitude, the amplitude of the oscillations at $x=0$ decays to zero as $\frac{1}{\sqrt{t}}$.  

{\em The NLSE and Rogue Waves in a $\phi^4$ Model.}
We next consider the $\phi^4$ theory given by the Lagrangian density
\begin{eqnarray}
\mathcal{L}=\frac{1}{2}\partial^\mu\phi^*\partial_\mu\phi-\frac{m^2}{2}|\phi|^2-\frac{\lambda}{4!}|\phi|^4,
\end{eqnarray}  
with $\phi$ a complex scalar field and implicit summation over space-time coordinates $\mu$.  The Euler-Lagrange equation, in the non-relativistic limit, is a NLSE with a potential shift, $m$:
\begin{eqnarray}
\label{potentialShift}
i\phi_t=-\frac{\nabla^2}{2m}\phi+\frac{\lambda}{12m}|\phi|^2\phi+m\phi
\end{eqnarray}
Solutions are given by those of Eq. (\ref{NLSE} ) with $F=|\Psi|^2$ multiplied by $e^{-i m t}$. If we discretize the $\phi^4$ Lagrangian density and consider a 2-dimensional lattice of points with 
spacing $\Delta$, the imaginary time $(t=i\tau)$ Lagrangian becomes
\begin{eqnarray}
\label{DiscreteLagrangian}
L=\displaystyle\sum\limits_{i,j}\left[\frac{1}{2}|\dot{\phi}_{ij}|^2+\frac{16+m^2\Delta^2}{2\Delta^2}|\phi_{ij}|^2+\frac{\lambda}{4!}|\phi_{ij}|^4\right]
\\\nonumber-\frac{1}{\Delta^2}\displaystyle\sum\limits_{\alpha}\delta_\alpha \phi_{ij}\phi_{kl}
\end{eqnarray}
 where the indices $i$, $j$ denote the two Cartesian coordinates of points in the plane, 
 $\alpha$ denotes each pair $(i,j),(k,l)$, and $\delta_\alpha$ is zero for non-adjacent and one for adjacent  oscillator pairs.

\begin{figure}[htbp!]
\centering
\subfigure[]{
\label{t=0}
\includegraphics[scale=.11]{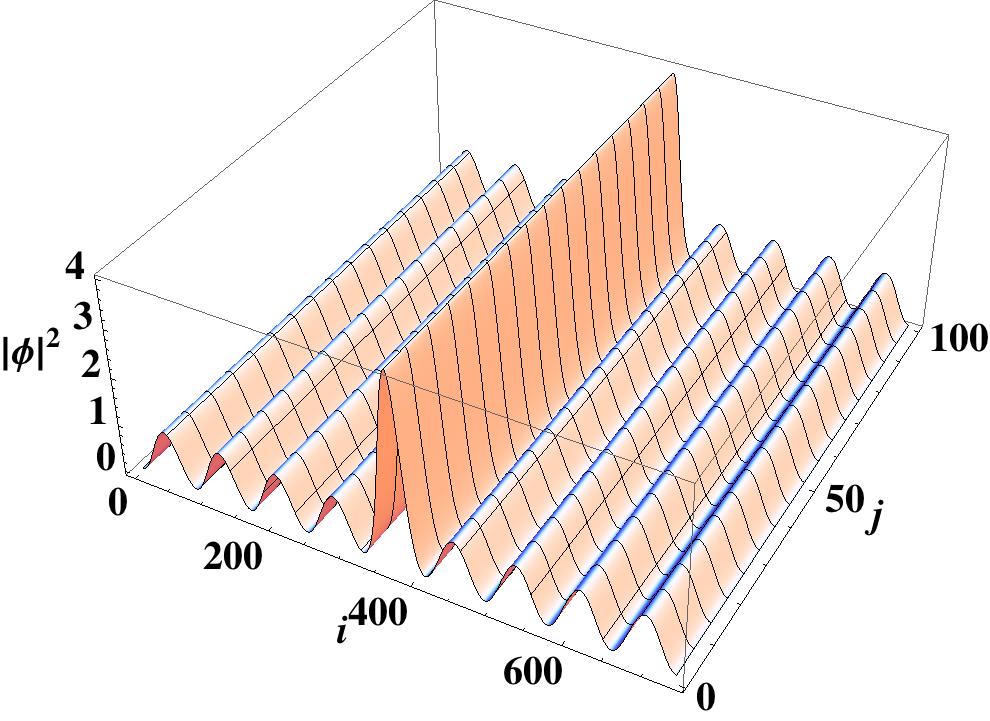}
}
\subfigure[]{
\label{t=60000}
\includegraphics[scale=.11]{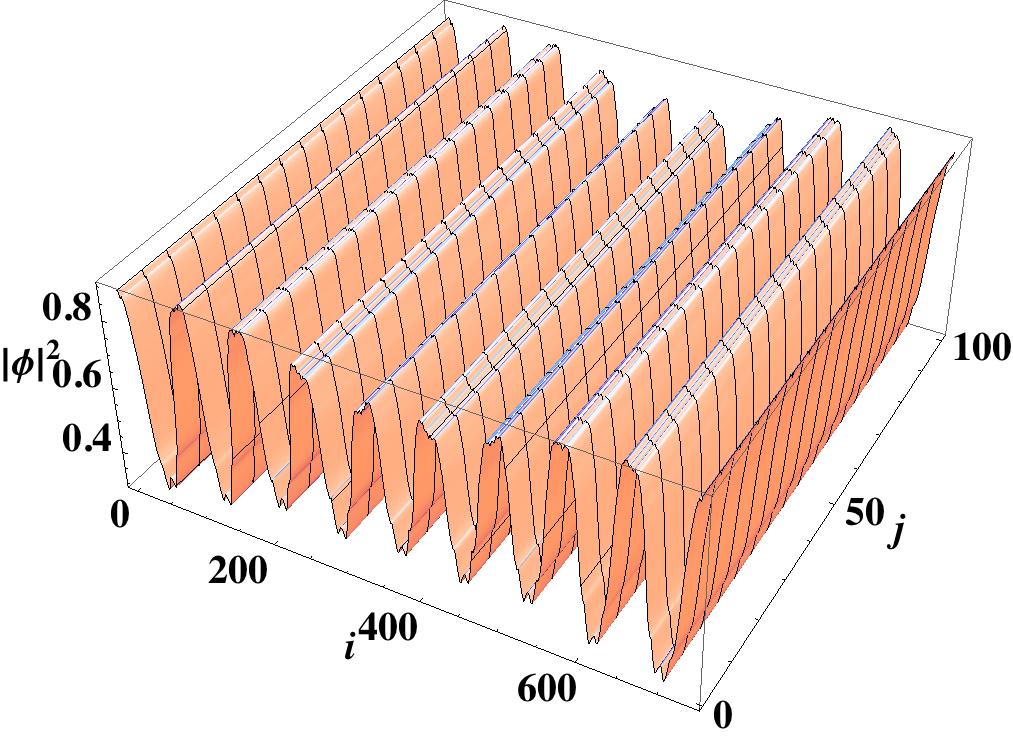}
}
\caption{(a) shows a region of the cell surrounding an initial Gaussian rogue wave on a cosine background where ($i$, $j$) denote Cartesian coordinates and $|\phi|^2$ denotes the modulus of the squared amplitude. (b) shows the same system at large times ($t=60,000$ iteration steps).}
\end{figure}

In the linear ($\lambda=0$) case, the system is quadratic and solvable by diagonalization.  We analyzed the time evolution for different parameters by a perturbative scheme in $\lambda$. In order to underscore the ubiquitous nature of rogue waves (including, as noted in some of our earlier
discussions (e.g., Eq.(\ref{roguewaveequation}), their presence already at the linear level), we plot below the evolution of an initial rogue wave state for the coupled linear oscillator Lagrangian of Eq. (\ref{DiscreteLagrangian}) with the parameters $m=2$ and $\Delta=1$.  The even appearance and low amplitude of the surface at large times illustrates that, already at a linear level 
(i.e., that with $\lambda =0$), the dispersion of the normal modes of the unperturbed sysetm ($\lambda =0$) enables an evolution of steady states into rogue wave configurations.  As the theory is \textit{time-reversal} invariant, we can conclude that a steady state given by Fig \ref{t=60000} could evolve into the rogue wave in Fig. \ref{t=0}.

{\em Conclusions.}
We investigated the NLSE by considering a corresponding classical
mechanics problem of non-linear oscillators. The focus
was on determining oscillatory waves with a general 
drift velocity $c$. By invoking this analogy, \newline
(1) We determined solutions to general NLSE with a general drift velocity $c$
in 1+1 dimensions for an arbitrary non-linearity
$F$ in Eq.(\ref{NLSE}) . The general result is that of Eq.(\ref{EllipticIntegral}). \newline
(2) We discussed solutions in general dimensions for large radial coordinates.  \newline
(3) We discussed the effects of noise and external sources. \newline
(4) We demonstrated that solutions could potentially be truncated and attached to each other
(as in Eq. (\ref{roguewaveequation}) in order to create approximate solutions that behave very much like a rogue wave.  We thus introduced the notion of  {\em ``generalized compactons''} which unlike usual compactons (and solitons) have a {\em non-zero (periodic) background}. The lifetime of
such solutions scales as the reciprocal of difference in currents in the vicinity of the stitching
points as given by Eq.(\ref{lifetime}).  We qualitatively saw the behavior of the amplitude of a rogue wave over time by solving the linear Schrodinger equation for rogue wave-like initial conditions.\newline
(5) We, specifically, employed considerations of the {\em continuity equation} to discuss life-times
of rogue wave type states.
(6) We reported on the appearance of rogue waves in $\phi^4$ theories. Specifically, we modeled rogue waves in a system of coupled oscillators.  We showed that this approach allows initial rogue wave-like conditions to evolve into what appears to be a steady state.  By {\em time-reversal invariance} of the equations of motion, we can conclude that rogue waves can evolve out of a steady state of that form.

\bibliography{PaperDraftBib}

\end{document}